\documentclass[prb,twocolumn,showpacs,preprintnumbers]{revtex4}

\usepackage{graphicx}
\usepackage{dcolumn}
\usepackage{amsmath}
\usepackage{color}

\begin{document}

\title{Pressure-induced spin-state transition of iron in magnesiow\"ustite (Fe,Mg)O}

\author{I. Leonov}
\affiliation{Theoretical Physics III, Center for Electronic Correlations and
Magnetism, Institute of Physics, University of Augsburg, 86135 Augsburg, Germany}
\affiliation{Materials Modeling and Development Laboratory, National University of Science and Technology 'MISIS', 119049 Moscow, Russia}

\author{A. Ponomareva}
\affiliation{Materials Modeling and Development Laboratory, National University of Science and Technology 'MISIS', 119049 Moscow, Russia}

\author{R. Nazarov}
\affiliation{Physics Division, Physical and Life Sciences Directorate, Lawrence Livermore National Laboratory, Livermore, CA 94551}

\author{I. A. Abrikosov}
\affiliation{Materials Modeling and Development Laboratory, National University of Science and Technology 'MISIS', 119049 Moscow, Russia}
\affiliation{Department of Physics, Chemistry and Biology (IFM), Link\"oping University, SE-58183 Link\"oping, Sweden}

\date{\today}

\begin{abstract}
We present a detailed theoretical study of the electronic, magnetic, and structural properties of
magnesiow\"ustite Fe$_{1-x}$Mg$_x$O with $x$ in the range between 0$-$0.875 using a fully charge self-consistent implementation of the density functional theory plus dynamical mean-field theory (DFT+DMFT) method. In particular, we compute the electronic structure and phase stability of the rock-salt B1-structured (Fe,Mg)O at high pressures relevant for the Earth's lower mantle.
We obtain that upon compression paramagnetic (Fe,Mg)O exhibits a spin-state transition of Fe$^{2+}$ ions from a high-spin to low-spin (HS-LS) state which is accompanied by a collapse of local magnetic moments. 
The HS-LS transition results in a substantial drop of the lattice volume by about 4$-$8 \%,
implying a complex interplay between electronic and lattice degrees of freedom.
Our results reveal a strong sensitivity of the calculated transition pressure $P_{\rm tr.}$ upon addition of Mg. While for Fe-rich magnesiow\"ustite, Mg $x < 0.5$, $P_{\rm tr.}$ exhibits a rather weak variation at $\sim$80 GPa, for Fe-poor (Fe,Mg)O it drops, e.g., by about 35 \% to 52 GPa for Mg $x=0.75$.
This behavior is accompanied by a substantial change of the spin transition range from 50$-$140 GPa in FeO to 30$-$90 GPa for $x=0.75$. In addition, the calculated bulk modulus (in the HS state) is found to increase by $\sim$12 \% from 142 GPa in FeO to 159 GPa in (Fe,Mg)O with Mg $x=0.875$.
We find that the pressure-induced HS-LS transition has different consequences for the electronic properties of
the Fe-rich and poor (Fe,Mg)O. For the Fe-rich (Fe,Mg)O, the transition is found to be accompanied
by a Mott insulator to (semi-) metal phase transition. In contrast to that, for $x>0.25$, (Fe,Mg)O remains insulating up to the highest studied pressures, implying a Mott insulator to band insulator phase transition at the HS-LS transformation.
%
%Our results for the electronic structure and lattice properties of FeO agree well with experimental data.

\end{abstract}

%\pacs{71.10.-w, 71.27.+a, 71.30.+h} 
\maketitle

%%%%%%%%%%%%%%%%%%%%%%%
\section{Introduction}
%%%%%%%%%%%%%%%%%%%%%%%

Magnesiow\"ustite (Fe$_{1-x}$,Mg$_x$)O is the second most abundant mineral in the Earth's interior which makes up some 20 \% of the total volume of Earth's lower mantle \cite{RevGeophys.51.244}. Therefore its high-pressure electronic properties, spin-state of iron, and phase stability play an important role in the (geo-) physics, chemistry and dynamics of the Earth's mantle. 
The high-pressure properties of (Fe,Mg)O have attracted much recent interest both from a theoretical and experimental point of views. At ambient conditions, (Fe,Mg)O is known to exists as a solid solution between  periclase (MgO) and w\"ustite (FeO). It has a rock-salt B1 crystal structure with Mg$^{2+}$ and high-spin (S=2) Fe$^{2+}$ ions having octahedral environments. Furthermore, (Fe,Mg)O is likely to keep the B1-type lattice structure throughout the Earth's lower mantle conditions as suggested by recent x-ray diffraction measurements \cite{PNAS.100.4405}. 
(Fe,Mg)O comprises two end-member oxides with remarkably different electronic properties: MgO and FeO. MgO is a band insulator with a B1-type crystal structure stable up to 227 GPa \cite{PhysRevLett.74.1371}, whereas FeO is a prototypical Mott insulator with a complex interplay between electronic structure and lattice under pressure \cite{JGRB.96.16169,PhysRevLett.79.5046,PhysRevLett.83.4101,PhysRevLett.93.215502,PhysEarthPlanInt.146.273,Science.266.1678,Science.334.792,GRL.38.24301,EarthPlanScL.304.496,PhysRevLett.108.026403}. By changing the Mg content $x$, it seems therefore become possible to tune a Mott to band insulator transition in (Fe,Mg)O \cite{PhysRevB.73.245118,PhysRevB.75.193103,PhysRevLett.99.126405,PhysRevB.80.155116,PhysRevB.89.035139}. 

High-pressure x-ray emission and M\"ossbauer spectroscopy experiments show that the Fe$^{2+}$ ion of FeO and (Fe,Mg)O undergoes a high-spin (HS) to low-spin (LS) transition, at pressures relevant for the Earth's lower mantle \cite{SSCom.59.513,PhysRevLett.79.5046,Science.300.789,Science.312.1205,Nature.436.377,GRL.34.23467,Science.317.1740,Science.324.224,PhysRevB.73.100101,GRL.38.24301,EarthPlanScL.304.496,JGRB.119.50699,GRL.38.L23308,AmMiner.101.1084,PNAS.102.17918,PNAS.100.4405,PNAS.110.7142}.
It has been confirmed that FeO makes a Mott insulator-to-metal transition at about 70 GPa, retaining the B1-type lattice structure at high temperature \cite{PhysRevLett.108.026403,GRL.38.24301,EarthPlanScL.304.496}.
For (Fe,Mg)O, these studies reveal that the transition pressure decreases upon increase of the Mg content. 
They also indicate that the spin-pairing transition affects electronic and elastic properties of (Fe,Mg)O and therefore has significant implications for the physics and chemistry of Earth.
On the theoretical side, the electronic properties of FeO and (Fe,Mg)O have been intensively investigated employing band structure based techniques \cite{Science.275.654,PhysRevB.47.7720,PhysRevB.55.12822,AmMineral.88.257,JETP.94.192,JETP.96.129,PhysRevLett.96.198501,GRL.33.16306,PNAS.106.8447,PhysRevLett.110.228501,PhysRevLett.114.117202,PNAS.111.10468}. These studies confirm a remarkable composition dependence of the pressure induced spin-state transition of Fe$^{2+}$ in (Fe,Mg)O, showing however a broad scattering for the calculated transition pressures. Therefore the effects of temperature and composition on the spin-state transition pressure and broadness of the spin crossover remained uncertain. All this makes a detailed study of the entire solid solution of (Fe,Mg)O to be essential for understanding its electronic state and magnetic properties.

These experimental and theoretical studies have lead us to reinvestigate the properties of the B1-type (Fe,Mg)O at high pressures employing a fully charge self-consistent implementation of the density functional plus dynamical mean-field theory method (DFT+DMFT) of strongly correlated electrons \cite{PhysRevLett.62.324,RevModPhys.78.865,RevModPhys.68.13,PhysToday.57.53}.
The DFT+DMFT method \cite{PhysRevB.75.155113,PhysRevB.76.235101,PhysRevB.77.205112,PhysRevB.90.235103} allows one to capture all generic aspects of a pressure-induced Mott
insulator-to-metal phase transition (MIT), such as a coherent quasiparticle behavior, formation
of the lower- and upper-Hubbard bands, and strong renormalization of the
effective electron mass (reduced electron mobility) \cite{Science.410.793,PhysRevB.70.205116,NatMater.7.198,PhysRevLett.99.156404,PhysRevLett.102.146402,PhysRevLett.106.106405,JPhysCondMat.27.275501,PhysRevB.81.075109,PhysRevB.82.195101,PhysRevB.85.020401,PhysRevB.86.195104,PhysRevB.90.155120,PhysRevB.94.100102}. Most importantly, applications of DFT+DMFT have shown to provide a good qualitative and even quantitative description of the electronic structure and phase stability of correlated materials, even in the vicinity of a Mott MIT \cite{PhysRevB.86.155121,PhysRevB.91.195115,PhysRevB.92.085142,PhysRevB.94.155135}.

We employ DFT+DMFT to investigate the electronic structure, spin-state of iron, and phase stability of \emph{paramagnetic} (Fe,Mg)O at high pressure for a broad range of Mg compositions $x=0$$-$$0.875$, which remained unexplored up to now.
Our results reveal that (Fe,Mg)O exhibits a pressure-induced spin-state transition of Fe$^{2+}$ ions from the high-spin (HS) to low-spin (LS) state which is accompanied by a collapse of local moments. 
Our results show a strong sensitivity of the electronic and lattice properties, transition pressure and transition range of a HS-to-LS state crossover upon Mg content $x$, indicating a complex interplay between electronic and lattice degrees of freedom. 
For Fe-rich (Fe,Mg)O, the HS-LS transition is found to be accompanied by a Mott insulator to (semi-) metal phase transition. In contrast to that, for the Mg content above 0.25, 
(Fe,Mg)O remains insulating up to the highest studied pressures. This implies that the HS-LS transition is accompanied by a Mott insulator to band insulator phase transition for $x > 0.25$.
Our results for the electronic structure and lattice properties are in overall good agreement with experimental data.

%%%%%%%%%%%%%%%%%
\section{Method}
%%%%%%%%%%%%%%%%%

In this work, we provide a detailed theoretical study of the electronic structure, magnetic state, and phase stability of paramagnetic B1-structured (Fe,Mg)O using a fully charge self-consistent implementation of the DFT+DMFT method \cite{PhysRevB.75.155113,PhysRevB.76.235101,PhysRevB.77.205112,PhysRevB.90.235103,PhysRevB.91.195115,PhysRevB.92.085142,PhysRevB.94.155135}. We use this advanced theory to compute the high pressure and temperature properties of (Fe,Mg)O as a function of Mg content $x$ in the range between 0$-$0.875, i.e., above the percolation limit ($\sim$12 \% Fe) of the face-centered cubic lattice of B1-type (Fe,Mg)O \cite{PhysRevE.57.230}. To this end, we calculate the total energy and (instantaneous) local magnetic moments $\sqrt{\langle m_z^2 \rangle}$ of B1-type (Fe,Mg)O as a function of lattice volume for different Mg $x$ \footnote{In our calculations, we do take take into account a possible decomposition of (Fe,Mg)O under high-pressure high-temperature conditions, as proposed in Ref.~\onlinecite{Science.289.430}.}. To model a chemical substitution Fe/Mg, we construct a supercell (with periodic boundary conditions) containing eight formula units of the host material FeO in which one to seven Fe ions were replaced with Mg. 
The positions of the impurity atoms (Mg/Fe) were arranged to maximize a distance from each other \footnote{We note that by construction this procedure preserves the fcc symmetry of the unitcell.} (Fe/Mg atoms are uniformly distributed over the unit cell, i.e, we neglect possible formation of the Fe/Mg clusters under pressure \cite{PhysRevB.80.014204}). For simplicity, we neglect the local relaxation effects around the impurity Mg/Fe atoms, as well as a possible formation of a site-selective Mott insulating phase with coexisting (within a unit cell) the HS and LS iron sites \cite{Greenberg_2017}. 
In order to evaluate pressure, we fit our total-energy results to the third-order Birch-Murnaghan equation of states \cite{PhysRev.71.809}, separately for the HS and the LS volume regions. The compressed phase is denoted by the relative volume w.r.t. the calculated equilibrium lattice volume as $\nu \equiv V/V_0$.

We employ the DFT+DMFT approach implemented within the plane-wave pseudopotentials 
\cite{PhysRevB.91.195115,PhysRevB.92.085142,PhysRevB.94.155135} with the generalized gradient approximation in DFT \cite{RevModPhys.73.515,JPhysCondMat.21.395502}. For the partially filled Fe $3d$ and O $2p$ orbitals we construct a basis set of Wannier functions \cite{PhysRevB.56.12847,RevModPhys.84.1419} using the projection procedure onto a local atomic-centered symmetry-constrained basis set as discussed in 
Refs.~\onlinecite{PhysRevB.71.125119,JPhysCondMat.20.135227,EurPhysJB.65.91}, with a window spanning both the Fe $3d$ and O $2p$ bands.
We model a chemical disorder in (Fe,Mg)O by applying averaging of the Green's functions of the Fe sites in accord with coherent potential approximation \cite{RPP.71.046501}. We employ a single-site DFT+DMFT approach to treat the effects of electron correlation in the Fe $3d$ shell, i.e., neglect the effect of spatial (non-local) correlations. 
To solve the 
realistic many-body problem, we employ the continuous-time hybridization-expansion (segment)
quantum Monte-Carlo algorithm \cite{RevModPhys.83.349}. The calculations are performed in the 
paramagnetic state at an electronic temperature $T = 1160$ K. In accordance with previous studies 
of FeO, we use the local Coulomb interaction $U=7$ eV and Hund's exchange 
$J=0.89$ eV parameters for the Fe $3d$ orbitals \cite{PhysRevB.82.195101,PhysRevLett.108.026403,PhysRevB.92.085142,JPhysCondMat.27.275501,PhysRevB.94.155135}. The $U$ and $J$ 
values are assumed to remain constant upon variation of the lattice volume. The Coulomb 
interaction is treated in the density-density approximation. The spin-orbit 
coupling is neglected in these calculations. We employ the 
fully-localized double-counting correction, evaluated from the self-consistently 
determined local occupations, to account for the electronic interactions already 
described by DFT. The spectral functions were computed using the maximum entropy 
method. The angle resolved spectra were evaluated from analytic continuation of 
the self-energy using Pad\'e approximants.

\section{Results and discussion}

%%%%%%%%%%%%%%%%%%%%%%%%%%%%%%%%%%%%%%
% DFT+DMFT calcululations of pure FeO
%%%%%%%%%%%%%%%%%%%%%%%%%%%%%%%%%%%%%%

%%%%%%%%%%%%%%%%%%%%%%%%%%%%%%%%%%%%%%%%%%%%%%%%%%%%%%%%%%%%%%%%%%%%%%%%%%%%%%%%%
\begin{figure}[tbp!]
\centerline{\includegraphics[width=0.5\textwidth,clip=true]{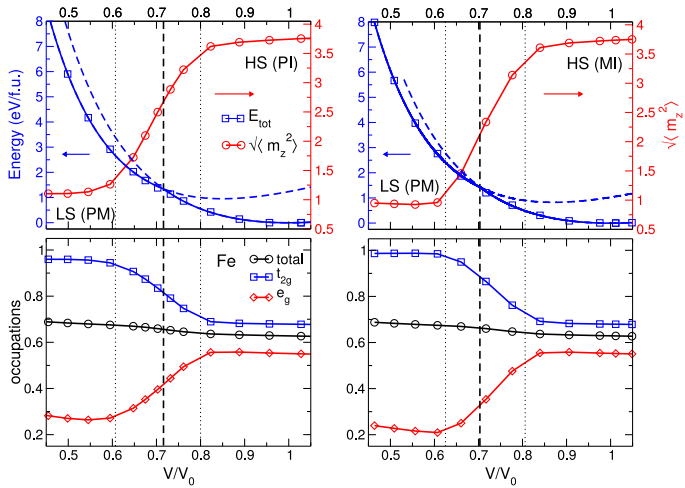}}
\caption{DFT+DMFT results for the total energy (blue) and instantaneous magnetic moments $\sqrt{\langle m_z^2 \rangle}$ (red) of paramagnetic FeO (left) and (Fe$_{0.75}$,Mg$_{0.25}$)O (right) as a function of
lattice volume. The HS-LS state transition is depicted by a vertical black dashed line. The HS-LS transition range (upper and lower onset of the transition) is shown by two vertical dotted lines. Bottom: Fe $3d$ and partial $t_{2g}$/$e_g$ occupations as a function of volume.
}
\label{Fig_1}
\end{figure}
%%%%%%%%%%%%%%%%%%%%%%%%%%%%%%%%%%%%%%%%%%%%%%%%%%%%%%%%%%%%%%%%%%%%%%%%%%%%%%%%%

As a starting point, we calculate the electronic structure, magnetic state, and phase stability of the eighth-formula-units supercell of the B1-structured paramagnetic phase of pure FeO (Mg $x=0$). In Fig.~\ref{Fig_1} (left) we display our results for the total energy and (instantaneous) magnetic local moments 
$\sqrt{\langle m_z^2 \rangle}$ computed within DFT+DMFT for different compression of the lattice ($\nu \equiv V/V_0$).
Our results agree quantitatively well with those previously published in Refs.~\onlinecite{PhysRevB.92.085142,PhysRevB.94.155135}. In particular, within the B1 lattice structure of FeO, a high-spin to low-spin transition is found to occur upon compression above $\sim$73 GPa.
The calculated bulk modulus $K_{0,T}$ for the low-pressure phase is 142 GPa, the (instantaneous) local magnetic moment $\sqrt{\langle m_z^2 \rangle} \sim 3.7$ $\mu_B$ that corresponds to a fluctuating moment of $\sim$3.6 $\mu_B$. Our results show that the bulk modulus in the LS phase of FeO is substantially larger than that in the HS phase (142 GPa). In fact, for the LS state our estimate of $K_{0,T}$ is about 210 GPa \footnote{
Interestingly, that for the LS state the estimated value of $K_{0,T}$ depends very sensitively on the details of the total-energy fitting, e.g., on the choice of the LS region. This can lead to a sufficient underestimation of $K_{0,T}$, e.g., in the previous reports \cite{PhysRevB.92.085142,PhysRevB.94.155135} it is suggested that $K_{0,T} \sim 162$ GPa.}.
The HS-LS state transformation is accompanied by a Mott insulator-to-metal phase transition \cite{PhysRevB.82.195101} with a drop of the lattice volume by about 9 \% at the MIT, implying a complex interplay between electronic and lattice degrees of freedom \cite{PhysRevB.92.085142,PhysRevB.94.155135}. Under pressure, our results indicate a substantial charge transfer in the Fe$^{2+}$ $3d$ shell between the $t_{2g}$ and $e_g$ states. Namely, the occupancy of the $t_{2g}$ orbitals gradually increases, resulting in (almost) completely occupied state (with the $t_{2g}$ occupation of about 0.95). Contrary to that, the $e_g$ orbitals are strongly depopulated (their occupation is below 0.3) while the total Fe $3d$ occupancy remains essentially unchanged with pressure. Our results for the high-pressure electronic, magnetic, and lattice properties of FeO, e.g., that above $\sim$73 GPa the B1-structured FeO undergoes a HS-to-LS transition that is accompanied by a Mott MIT and collapse of the lattice volume, are in overall good agreement with recent experimental data \cite{GRL.38.24301,EarthPlanScL.304.496,PhysRevLett.108.026403}. Moreover, in accordance with previous studies, our calculations clearly indicate the crucial importance of electronic correlations for the high-pressure properties of FeO \cite{PhysRevLett.108.026403,PhysRevB.92.085142,PhysRevB.94.155135}.

%%%%%%%%%%%%%%%%%%%%%%%%%%%%%%%%%%%%%%%%%%%%%%%%%%%%%%%%%%%%%%%%%%%%%%%%%%%%%%%%%
\begin{figure}[tbp!]
\centerline{\includegraphics[width=0.5\textwidth,clip=true]{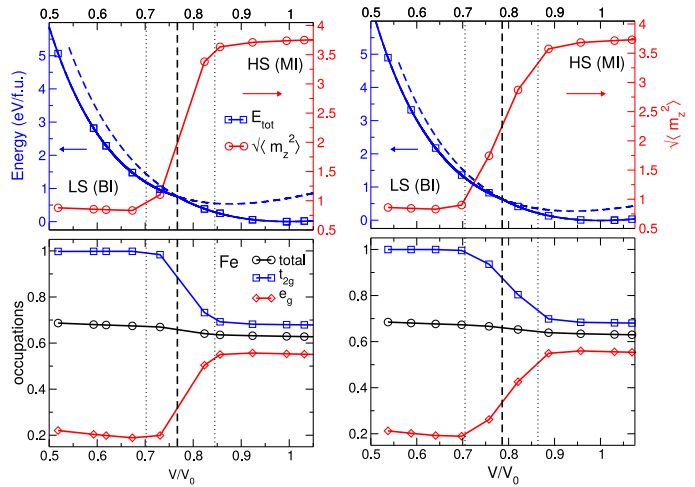}}
\caption{Our results for total energy (red) and instantaneous magnetic moments $\sqrt{\langle m_z^2 \rangle}$ (blue) of magnesiow\"ustite with Mg content $x=0.5$ (left) and 0.75 (right) calculated by DFT+DMFT for different lattice volumes. The Fe $3d$ and partial $t_{2g}$/$e_g$ occupations as a function of volume are shown in bottom.
}
\label{Fig_2}
\end{figure}
%%%%%%%%%%%%%%%%%%%%%%%%%%%%%%%%%%%%%%%%%%%%%%%%%%%%%%%%%%%%%%%%%%%%%%%%%%%%%%%%%

%%%%%%%%%%%%%%%%%%%%%%%%%%%%%%%%%%%%%%%%%%%%%%%%%%%
% DFT+DMFT total energy calcululations of (Fe,Mg)O
%%%%%%%%%%%%%%%%%%%%%%%%%%%%%%%%%%%%%%%%%%%%%%%%%%%

\begin{figure}[tbp!]
\centerline{\includegraphics[width=0.45\textwidth,clip=true]{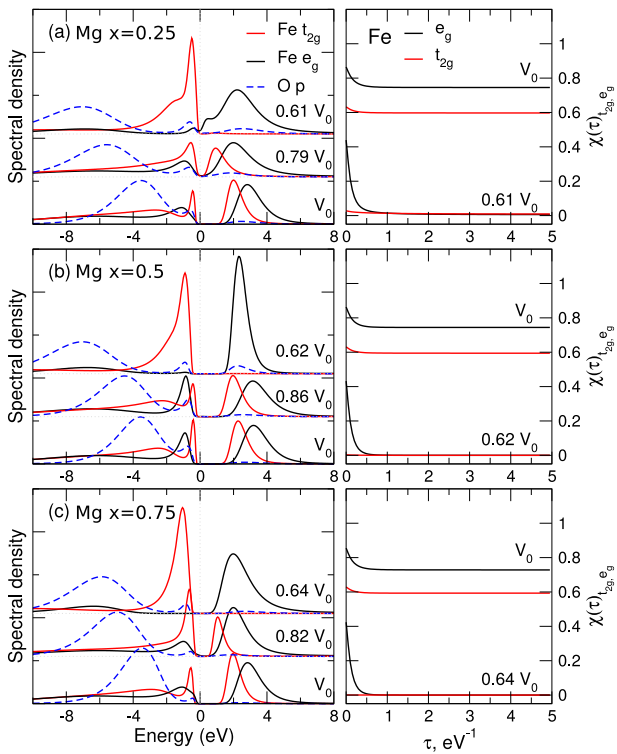}}
\caption{Left panel: Partial Fe $t_{2g}$/$e_g$ and O $2p$ spectral functions of magnesiow\"ustite with Mg content $x = 0.25$ (a), 0.5 (b), and 0.75 (c) calculated by DFT+DMFT for different lattice volumes. Right panel: Local spin-spin correlation function $\chi(\tau)$ calculated by DFT+DMFT as a function of volume. The intra-orbital $t_{2g}$ and $e_g$ contributions are shown.
}
\label{Fig_3}
\end{figure}

Next we compute the electronic structure and lattice properties of (Fe,Mg)O as a function of Mg content $x$ under pressures relevant to the Earth's lower mantle conditions. In Figs.~\ref{Fig_1} and \ref{Fig_2} we present our results for the total energy and local moments $\sqrt{\langle m_z^2 \rangle}$ of the B1-structured (Fe,Mg)O calculated within DFT+DMFT for different $\nu \equiv V/V_0$. Our results for the bulk modulus and equilibrium lattice volume evaluated from the DFT+DMFT total-energy calculations are summarized in Table~\ref{tab:Table1}. At ambient pressure, for all $x$ we obtain a Mott insulating solution with a large $d$-$d$ energy gap of about 2 eV, in accordance with with previous studies \cite{JPhysCondMat.27.275501}. Our results for the Fe $t_{2g}$ and $e_g$ orbital occupations are about 0.65 and 0.55, respectively, near the half-filling, implying the HS (S=2) state of Fe$^{2+}$ ions. In addition, similar to FeO, the instantaneous local moment is about 3.7 $\mu_B$ (fluctuating moment of 3.6 $\mu_B$). The Fe $3d$ electrons are localized as seen from our result for the local spin-spin correlation function $\chi(\tau)=\langle \hat{m}(\tau) \hat{m}(0) \rangle$ shown in Fig. \ref{Fig_3} (where $\tau$ is the imaginary time). In fact, $\chi(\tau)$ is seen to be almost
constant and close to its maximal value for the partial Fe $3d$ states (i.e., to unity), indicating localization of $3d$ electrons at ambient pressure. We also point out the crucial importance of the effects of electron correlation to determine the electronic properties of (Fe,Mg)O.

\begin{table*}[t]
\caption{Calculated structural parameters for the paramagnetic $B1$ phase of (Fe,Mg)O for different Mg content $x$. $V_0$ is ambient pressure volume. $K_{0,T}$ is bulk modulus for the HS and LS phase; $K' \equiv dK/dP$ is fixed to 4.1 for all Mg compositions. $P_{\rm tr.}$ is the HS-LS transition pressure. $|\Delta V|$ and $\Delta V/V$ is an absolute and relative change of the lattice volume at the HS-LS transition.}

\centering
\begin{tabular}{lcccccc} 
\hline
\hline
Mg$_x$ & $V_0$ (a.u.$^3$/f.u.) & $K_{0,T}^{HS}$ (GPa) & $K_{0,T}^{LS}$ (GPa) & $P_{\rm tr.}$ (GPa) & $|\Delta V|$ (a.u.$^3$/f.u.) & $\Delta V/V$ (\%) \\
\hline
0     & 144.1 & 142 & 210 & 73 & 10.2 & 9 \\
0.125 & 143.1 & 139 & 205 & 82 & 8.3  & 8 \\
0.25  & 141.3 & 137 & 201 & 83 & 7.1  & 7 \\
0.375 & 139.5 & 138 & 213 & 77 & 7.2  & 7 \\
0.5   & 138.6 & 139 & 200 & 49 & 8.6  & 8 \\
0.625 & 135.5 & 142 & 185 & 61 & 5.2  & 5 \\
0.75  & 133.8 & 151 & 169 & 52 & 4.7  & 4 \\
0.875 & 132.9 & 159 & 158 & 21 & 5.1  & 4 \\
\hline
\hline
\end{tabular}

\label{tab:Table1}
\end{table*}

%%%%%%%%%%%%%%%%%%%%%%%%%%%%%%%%%%%%%%%%%%%%%%%%%%%%%%%%%%%%%%%%%%%%%%%%%%%%%%%%%
%%%%%%%%%%%%%%%%%%%%%%%%%%%%%%%%%%%%%%%%%%%%%%%%%%%%%%%%%%%%%%%%%%%%%%%%%%%%%%%%%

%%%%%%%%%%%%%%%%%%%%%%%%%%%%%%%%%%%%%%%%%%%%%%%%%%%%%%%%%%%%%%%%%%%%%%%%%%%%%%%%%%%%%%
% DFT+DMFT spectral functions of (Fe,Mg)O: metalization vs. band insulating behavior 
%%%%%%%%%%%%%%%%%%%%%%%%%%%%%%%%%%%%%%%%%%%%%%%%%%%%%%%%%%%%%%%%%%%%%%%%%%%%%%%%%%%%%%

Upon compression, our calculations show that (Fe,Mg)O compounds undergo a HS-LS phase transition, with a collapse of the local moments to a LS state \cite{JPhysCondMat.27.275501}. The LS state is characterized by a fluctuating magnetic moment which is below $\sim$0.2$-$0.4 $\mu_B$ for pressures above $\sim$150 GPa, i.e., for $\nu \leq 0.6-0.7$. Interestingly, that at the same pressure, the LS FeO has a fluctuating moment of $\sim$0.7 $\mu_B$, i.e., remarkably higher than that in the LS state of (Fe,Mg)O. Similarly to FeO, we observe a substantial redistribution of charge between the Fe $t_{2g}$ and $e_g$ orbitals within the Fe $3d$ shell caused by applied pressure. Above the HS-LS transition, it leads to a (almost) complete occupation of the Fe $t_{2g}$ states, while the Fe $e_g$ states are strongly depopulated (with occupancy below 0.2$-$0.3). 

\begin{figure}[tbp!]
\centerline{\includegraphics[width=0.375\textwidth,clip=true]{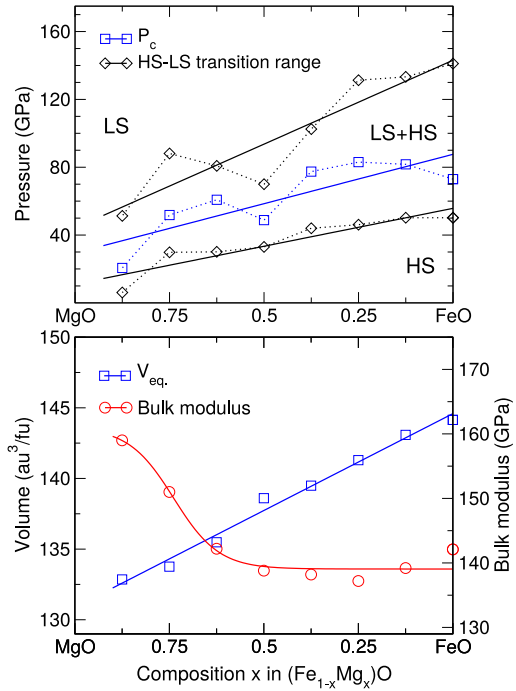}}
\caption{Our results for the HS-LS transition pressure and transition pressure range (upper and lower onset of the transition) [top] and the equilibrium lattice volume and bulk modulus [bottom] of (Fe,Mg)O as a function of Mg $x$ calculated by DFT+DMFT at $T=1160$ K.
}
\label{Fig_5}
\end{figure}

The HS-LS spin-state transition is found to be accompanied by a substantial drop of lattice volume of $\sim$4$-$8 \% (see Table~\ref{tab:Table1}). 
We note however that these values should be considered as an upper-bound estimate because we neglect multiple intermediate-phase transitions when fit the total-energy result to the third order Birch-Murnaghan equation of states \cite{PhysRev.71.809}.
The structural change takes place upon a compression of the lattice volume to $\nu \sim 0.7$$-$$0.8$. Our results for the calculated transition pressures are about 73 and 52 GPa for the Mg content of $x=0.25$ and 0.75, respectively. This implies that the electronic and structural properties of (Fe,Mg)O are strongly sensitive to addition of Mg. While for Fe-rich (Fe,Mg)O, for $x < 0.5$, the calculated transition pressure exhibits a rather weak variation at around 80 GPa, for the Fe-poor compounds the HS-LS transition pressure drops substantially, e.g., to 52 GPa, i.e., by $\sim$35 \%, for $x=0.75$. 
We also note a substantial increase from $\sim$140 to 160 GPa, i.e., by about 12 \%, of the calculated bulk modulus in HS (Fe,Mg)O for $x > 0.5$. This behavior is accompanied by a gradual decrease of the equilibrium lattice volume of (Fe,Mg)O as show in Fig. \ref{Fig_5}. In addition, we obtain a substantial change of the HS-LS transition range, from $\sim$50$-$140 GPa in FeO to 30$-$88 GPa in (Fe,Mg)O with Mg $x=0.75$. This indicates that the HS-LS transition width decreases with Mg $x$, in agreement with recent experiments \cite{PhysRevB.73.113107,AmMiner.101.1084}. 
%This behavior can be interpreted as a coexistence region of the HS and LS state of the Fe$^{2+}$ ions, suggesting the existence of a mixed-spin state (presumably due to strong spin fluctuations) with coexisting high-spin and low-spin iron states \cite{PhysRevLett.114.117202,JPhysCondMat.27.275501}.

%%%%%%%%%%%%%%%%%%%%%%%%%%%%%%%%%%%%%%%%%%%%%%%%%%%%%%%%%%%%%%%%%%%%%%%%%%%%%%%%%
\begin{figure}[tbp!]
\centerline{\includegraphics[width=0.35\textwidth,clip=true]{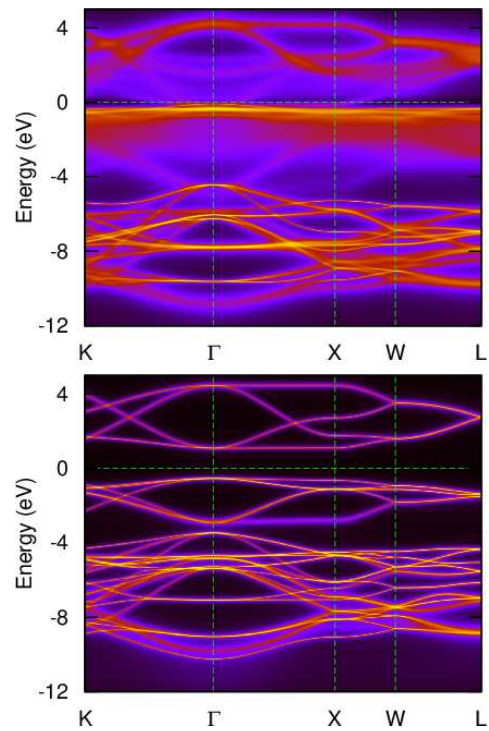}}
\caption{{\bf k}-resolved Fe $3d$ and O $2p$ spectral function of paramagnetic high-pressure phases of (Fe,Mg)O as obtained by DFT+DMFT. Top: Results for the LS phase of (Fe,Mg)O with Mg content $x=0.25$ and lattice volume $\nu \sim 0.61$. Bottom: Our results for the LS phase with $x=0.75$ and lattice volume compression of $\sim$0.64. 
}
\label{Fig_4}
\end{figure}
%%%%%%%%%%%%%%%%%%%%%%%%%%%%%%%%%%%%%%%%%%%%%%%%%%%%%%%%%%%%%%%%%%%%%%%%%%%%%%%%%

%%%%%%%%%%%%%%%%%%%%%%%%%%%%%%%%%%%%%%%%%%%%%%%%%%%%%%%%%%%%%%%%%%%%%%%%%%%
% DFT+DMFT spectral functions of (Fe,Mg)O: delocalization of 3d electrons
%%%%%%%%%%%%%%%%%%%%%%%%%%%%%%%%%%%%%%%%%%%%%%%%%%%%%%%%%%%%%%%%%%%%%%%%%%%

Our results for the electronic properties, equilibrium volume, and phase stability of (Fe,Mg)O with Mg $x=0$$-$$0.875$ calculated within DFT+DMFT agree well with available experimental data \cite{PNAS.102.17918,Nature.436.377,PhysRevB.73.113107,JGRB.119.50699,AmMiner.101.1084}. Overall, they are (qualitatively) consistent with the high-pressure behavior of pure FeO. Moreover, we observe a substantial change in the behavior of the Fe $3d$ electrons, which exhibits a crossover from a localized to itinerant magnetic behavior under pressure, implying delocalization of $3d$ electrons \cite{PhysRevB.92.085142,PhysRevB.94.155135}. Interestingly that similar to FeO, the calculated bulk moduli of (Fe,Mg)O exhibit a sharp increase at the HS-LS transition, except for Mg $x=0.75$, where $K_{0,T}$ remains essentially unchanged ($\sim$158-159 GPa) at the HS-LS transition. 
Furthermore we note that the Fe-rich
and poor (Fe,Mg)O, while both exhibit a HS-LS transition, show remarkably different electronic properties
at high pressures. In particular, for the Fe-rich (Fe,Mg)O compounds with Mg content $x \leq 0.25$, a HS-LS transition in the B1-type structure results in metallization, i.e., a Mott insulator to (semi-) metal phase transition. In fact, (Fe,Mg)O with $x=0.25$ shows a bad metal behavior at high pressures as shown in Fig. \ref{Fig_3} (a). In addition, our results for the {\bf k}-resolved spectral function of (Fe,Mg)O with $x=0.25$ (see Fig.~\ref{Fig_4}) show a semi-metallic behavior with a substantial broadening of the electronic states near the Fermi level due to the effect of electron-electron correlations.
In contrast to that, for the Fe-poor (Fe,Mg)O with Mg $x > 0.25$ the high-pressure LS phase is an insulator. Moreover, for magnesiow\"ustite with Mg $x > 0.25$, the energy gap (as it is partly seen in Fig.~\ref{Fig_3}) is found to increase upon compression above the HS-LS transition. Our analysis of the high-pressure behavior of the self-energy of the Fe-poor (Fe,Mg)O compounds suggest that the spin-pairing transition is accompanied by a Mott
insulator to band insulator phase transition \cite{PhysRevB.89.035139,PhysRevB.80.155116,PhysRevLett.99.126405,PhysRevB.75.193103}. 
Indeed, in the latter case, e.g., for (Fe,Mg)O with $x = 0.75$, the electronic states are seen to be highly coherent, revealing no finite-time broadening effects in electronic spectrum as usually caused by the effects of electron-electron correlations.
%It is seen as a high coherence of the electronic states of (Fe,Mg)O with $x = 0.75$ which reveal no finite-time broadening effects in electronic spectrum caused by the effects of electron-electron correlations.%
This implies that the effects of dynamical electronic correlations are weak for the Fe-poor (Fe,Mg)O, suggesting the importance of the effects of crystal-field splitting and their enhancement caused by static correlations.
%
%Moreover, the HS-LS transition results in a collapse of the lattice volume by $\sim 3$$-$$7$ \%.

%%%%%%%%%%%%%%%%%%%%%%%%%%%%%%%%%%%%%%%%%%%%%%%%%%%%%%%%%%%%%%%%%%%%%%%%%%%%%%%%%
%%%%%%%%%%%%%%%%%%%%%%%%%%%%%%%%%%%%%%%%%%%%%%%%%%%%%%%%%%%%%%%%%%%%%%%%%%%%%%%%%

%%%%%%%%%%%%%%%%%%%%%%%%%%%%%%%%%%%%%%%%%%%%%%%%%%%%%%%%%%%%%%%%%%%%%%%%%%
% (Fe,Mg)O transition pressure, bulk modulus and V_eq. for different Mg x
%%%%%%%%%%%%%%%%%%%%%%%%%%%%%%%%%%%%%%%%%%%%%%%%%%%%%%%%%%%%%%%%%%%%%%%%%%

In Fig. \ref{Fig_5} we summarize our results for the behavior of (Fe,Mg)O as a function of Mg content $x$ calculated within DFT+DMFT. We observe that (Fe,Mg)O compounds show a HS-LS phase transition, with a collapse of the local magnetic moment to a LS state. We obtain that the HS-LS transition pressure decreases upon addition of Mg, in agreement with experimental data \cite{PNAS.102.17918,Nature.436.377,AmMiner.101.1084}. This implies that addition of FeO in MgO results in stabilization the HS state of Fe$^{2+}$ to higher pressures. This behavior is accompanied with a substantial increase of the equilibrium volume of the HS phase of (Fe,Mg)O by $\sim$9 \% upon moving from MgO to FeO. 
For Fe-rich (Fe,Mg)O, the ambient-pressure bulk modulus $K_{0,T}$ shows a rather weak variation (about 140 GPa) with Mg $x$. For the Fe-poor compounds, $K_{0,T}$ is found to increase substantially by $\sim$12 \% for Mg $x > 0.5$.
We note that addition of Mg can be interpreted as an effective chemical
pressure acting on the high-spin Fe$^{2+}$ ion. This leads to a change of the effective Coulomb interaction strength to bandwidth ratio of iron ion in (Fe,Mg)O (here we do not consider the effects of percolation since our calculations were performed above the percolation limit of $\sim$12 \% Fe). This may result in a HS Mott insulator to LS band insulator
phase transition without metallization upon increasing of the crystal field
splitting caused by applied pressures, in qualitative agreement with the generic phase diagram 
of a HS-LS transition in the two-orbital Hubbard model \cite{PhysRevLett.106.256401}. Our results for the B1-structured
(Fe,Mg)O provide a unified picture of the HS-LS transition in magnesiow\"ustite. While
the Fe-rich (Fe,Mg)O exhibit a rather weak variation of the electronic structure and
lattice properties, the properties of the Fe-poor compounds are remarkably different. 
It appears to be due to a more local nature of magnetic interactions of Fe$^{2+}$ ion in the Fe-poor
compounds. Indeed, the contribution of the Fe-Fe exchange interaction which tends to
stabilize the HS state to much higher pressures is much weaker (or even absent) in Fe-poor
(Fe,Mg)O. This suggests the importance of percolation effects for understanding the properties of Fe-poor (Fe,Mg)O.

%%%%%%%%%%%%%%%
% Conclusions
%%%%%%%%%%%%%%%

In conclusion, we have determined the electronic properties, magnetic state, and phase stability of paramagnetic B1-structured magnesiow\"ustite Fe$_{1-x}$Mg$_x$O for Mg content $x$ in the range between 0$-$0.875 using DFT+DMFT. We computed the electronic structure and phase stability of the rock-salt B1-structured (Fe,Mg)O at high pressures and temperatures relevant for the Earth's lower mantle conditions.
Upon compression paramagnetic (Fe,Mg)O exhibits a spin-pairing transition of Fe$^{2+}$ ions which is accompanied by a simultaneous collapse of local moments. 
Our results reveal a strong sensitivity of the calculated transition pressure $P_{\rm tr.}$ upon addition of Mg. 
While for Fe-rich magnesiow\"ustite, $P_{\rm tr.}$ has a rather weak variation and is about 80 GPa for $x < 0.5$, for the Fe-poor case it drops to about 21 GPa for $x=0.875$.
In agreement with experiment, this behavior is accompanied by an increase of the calculated bulk modulus in the HS phase of (Fe,Mg)O by about 12 \% for Mg $x>0.5$. Moreover, the equilibrium lattice volume of (Fe,Mg)O shows a substantial monotonous decrease with Mg $x$. This suggests that addition of Mg can be interpreted as an effective chemical pressure acting on the high-spin Fe$^{2+}$ ion. Moreover, we find that addition of FeO in MgO stabilizes the high-spin state to higher pressures.
For the Fe-rich (Fe,Mg)O, the transition is found to be accompanied
by a Mott insulator to (semi-) metal phase transition. In contrast to that, for the Mg content above $x>0.25$ 
(Fe,Mg)O remains insulating up to the highest studied pressures. Our results suggest that for $x>0.25$ the HS-LS transition is accompanied by a Mott insulator to band insulator phase transition.
The lattice volume is found to collapse by about 4$-$8 \% at the HS-LS transition, implying a complex interplay between electronic and lattice degrees of freedom. 
Our results suggest that for the Mg compositions relevant for the Earth's interior, i.e., Mg $x=0.7$$-$$0.9$, Fe$^{2+}$ ion of (Fe,Mg)O is in a LS state throughout most of the Earth's lower mantle \cite{PNAS.102.17918}.
We point out the importance of further theoretical and experimental investigations of the behavior
of (Fe,Mg)O at high pressures and temperatures, e.g., studying the effect of short-range ordering and formation of Fe clusters \cite{PhysRevB.80.014204} and possible decomposition of (Fe,Mg)O \cite{Science.289.430,JGRB.119.50699}, for a better understanding of the Earth's lower mantle and outer core.

\begin{acknowledgments}

We thank D. Vollhardt, G. Kh. Rozenberg, and L. Dubrovinsky for valuable discussions. I.L. acknowledges support by the Deutsche Forschungsgemeinschaft through Transregio TRR 80 and the Ministry of Education and Science of the Russian Federation in the framework of Increase Competitiveness Program of NUST “MISIS” (K3-2016-027), implemented by a governmental decree dated 16th of March 2013, N 211. A.P. is grateful to the Russian Foundation for Basic Researches (Grant No. 16-02-00797) for financial support. I.A.A. gratefully acknowledges the Swedish Research Council (VR) grant
No. 2015-04391 and the Swedish Government Strategic Research Area in Materials Science
on Functional Materials at Link¨oping University (Faculty Grant SFO-Mat-LiU No.
2009 00971).

\end{acknowledgments}

\begin{appendix}
\section*{Appendix}

Here we present our results for the electronic structure and phase stability of (Fe,Mg)O calculated by DFT+DMFT for a number
of intermediate Mg contents.
%
%%%%%%%%%%%%%%%%%%%%%%%%%%%%%%%%%%%%%%%%%%%%%%%%%%%%%%%%%%%%%%%%%%%%%%%%%%%%%%%%%
\begin{figure}[tbp!]
\centerline{\includegraphics[width=0.5\textwidth,clip=true]{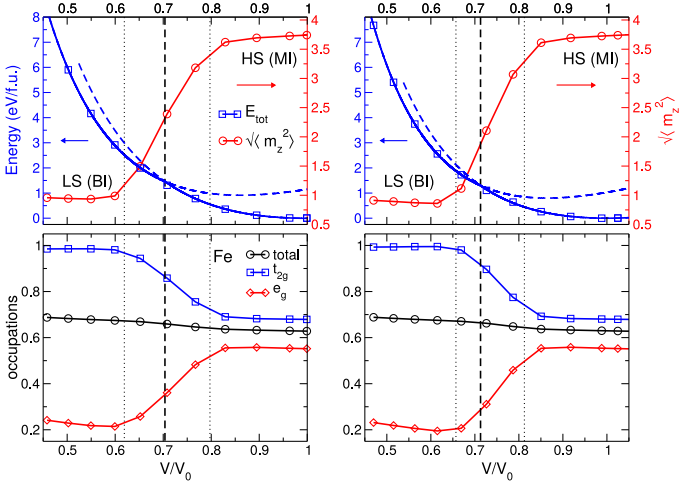}}
\caption{DFT+DMFT results for the total energy and instantaneous magnetic moments $\sqrt{\langle m_z^2 \rangle}$ of paramagnetic (Fe,Mg)O with Mg $x=0.125$ (left) and 0.375 (right) as a function of
volume. The HS-LS state transition is depicted by a vertical black dashed line. The HS-LS transition range is shown by two vertical dotted lines. Bottom: Fe $3d$ and partial $t_{2g}$/$e_g$ occupations as a function of volume.
}
\label{Fig_6}
\end{figure}
%%%%%%%%%%%%%%%%%%%%%%%%%%%%%%%%%%%%%%%%%%%%%%%%%%%%%%%%%%%%%%%%%%%%%%%%%%%%%%%%%
%
%%%%%%%%%%%%%%%%%%%%%%%%%%%%%%%%%%%%%%%%%%%%%%%%%%%%%%%%%%%%%%%%%%%%%%%%%%%%%%%%%
\begin{figure}[tbp!]
\centerline{\includegraphics[width=0.5\textwidth,clip=true]{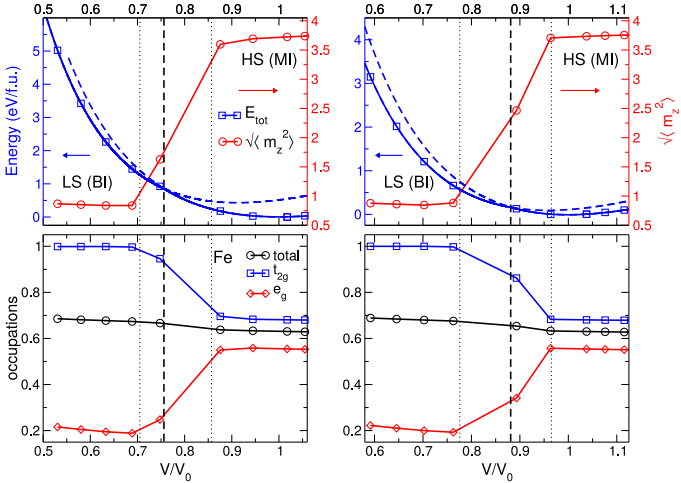}}
\caption{DFT+DMFT results as in Fig. \ref{Fig_6} for (Fe,Mg)O with Mg $x=0.625$ (left) and 0.875 (right).
}
\label{Fig_7}
\end{figure}
%%%%%%%%%%%%%%%%%%%%%%%%%%%%%%%%%%%%%%%%%%%%%%%%%%%%%%%%%%%%%%%%%%%%%%%%%%%%%%%%%

%%%%%%%%%%%%%%%%%%%%%%%%%%%%%%%%%%%%%%%%%%%%%%%%%%%%%%%%%%%%%%%%%%%%%%%%%%%%%%%%%
\begin{figure}[tbp!]
\centerline{\includegraphics[width=0.45\textwidth,clip=true]{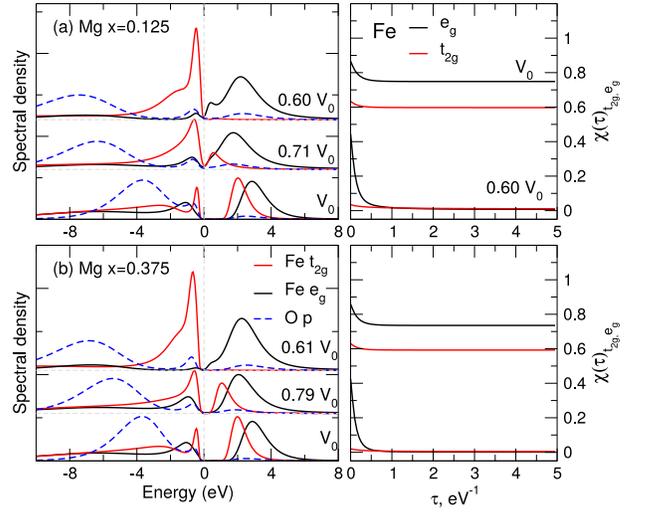}}
\caption{DFT+DMFT spectral functions of magnesiow\"ustite with Mg content $x = 0.125$ (a) and 0.375 (b) calculated by DFT+DMFT for different volumes. Partial Fe $t_{2g}$/$e_g$ and O $2p$ orbital contributions are presented. (Right panel:) Local spin-spin correlation function $\chi(\tau)$ calculated by DFT+DMFT as a function of volume. The intra-orbital $t_{2g}$ and $e_g$ contributions are shown.
}
\label{Fig_8}
\end{figure}
%%%%%%%%%%%%%%%%%%%%%%%%%%%%%%%%%%%%%%%%%%%%%%%%%%%%%%%%%%%%%%%%%%%%%%%%%%%%%%%%%
%
%%%%%%%%%%%%%%%%%%%%%%%%%%%%%%%%%%%%%%%%%%%%%%%%%%%%%%%%%%%%%%%%%%%%%%%%%%%%%%%%%
\begin{figure}[tbp!]
\centerline{\includegraphics[width=0.45\textwidth,clip=true]{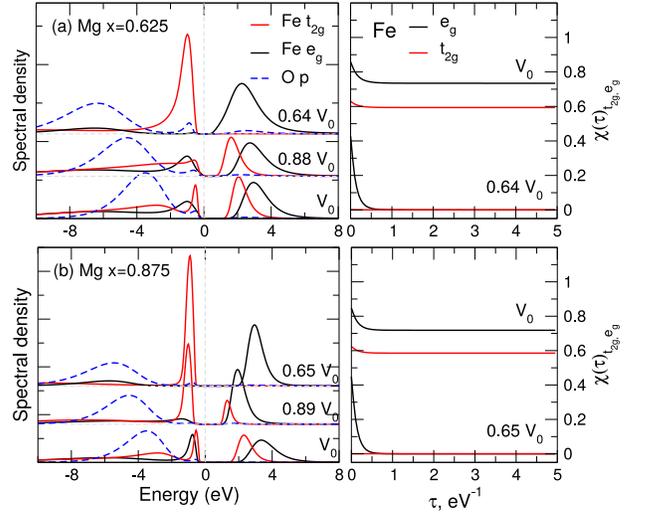}}
\caption{DFT+DMFT results as in Fig. \ref{Fig_8} for (Fe,Mg)O with Mg $x=0.625$ (a) and 0.875 (b).
}
\label{Fig_9}
\end{figure}
%%%%%%%%%%%%%%%%%%%%%%%%%%%%%%%%%%%%%%%%%%%%%%%%%%%%%%%%%%%%%%%%%%%%%%%%%%%%%%%%%

\end{appendix}

\bibliography{draft}

\end{document}